\documentclass[12pt]{article}
\usepackage[latin1]{inputenc}
\usepackage{graphicx}
\newcommand{\circrm}{{\rm circ}}

\newcommand{\text}{\rm}

\newcommand{\ug}{ \; = \; }

\newcommand{\infi}{\infty}

\newcommand{\bb}{\begin{equation}}
\newcommand{\ee}{\end{equation}}
\newcommand{\bega}{\begin{eqnarray}}
\newcommand{\ega}{\end{eqnarray}}
\newcommand{\begae}{\begin{eqnarray*}}
\newcommand{\egae}{\end{eqnarray*}}

\newcommand{\h}{\hspace*{4ex}}
\newcommand{\dis}{\displaystyle}

\newcommand{\om}{\omega}

\newcommand{\cent}{\centerline}
\newcommand{\vs}{\vspace*}

\newcommand{\kr}{k_{\rho}}

\begin{document}

\baselineskip 0.8cm

\begin{center}

{\large {\bf A simple and effective method for the {\em analytic} description of important optical beams, when truncated by
finite apertures} $^{\: (\dag)}$ }
\footnotetext{$^{\: (\dag)}$ The authors acknowledge partial support from FAPESP (under grant 11/51200-4);
from CNPq (under grant 307962/2010-5); as well as from CAPES and INFN. \
E-mail addresses for contacts: mzamboni@dmo.fee.unicamp.br; recami@mi.infn.it}


\end{center}

\vs{5mm}

\cent{ M. Zamboni-Rached, }

\vs{0.2 cm}

\centerline{{\em DMO--FEEC, University of Campinas, Campinas, SP, Brazil.}}

\vs{0.3 cm}

\cent{ Erasmo Recami }

\vs{0.2 cm}

\cent{{\em INFN---Sezione di Milano, Milan, Italy;}}

\cent{{\em Facolt\`a di Ingegneria, Universit\`a statale di Bergamo, Bergamo, Italy;}}

\cent{{{\rm and} \em DMO-FEEC, Universidade Estadual de Campinas, SP, Brazil.}}

\vs{0.2 cm}

\centerline{\rm and}

\vs{0.3 cm}

\cent{ Massimo Balma }

\vs{0.3 cm}

\cent{{\em SELEX Galileo, San Maurizio Can.se (TO), Italy.}}

\vs{0.5 cm}

{\bf Abstract  \ --} \ In this paper we present a simple and effective method, based on appropriate superpositions of Bessel-Gauss
beams, which in the Fresnel regime is able to
describe in {\em analytic} form the 3D evolution of important waves as Bessel beams, plane waves, gaussian beams, Bessel-Gauss
beams, when truncated by finite apertures. \ One of the
byproducts of our mathematical method is that one can get in few seconds, or minutes, high-precision results which normally
require quite long times of numerical simulation. \ The
method works in Electromagnetism (Optics, Microwaves,...), as well as in Acoustics.


\section{Introduction}

\h The analytic description of wave beam (in particular optical beam) propagation is of high importance, both in theory and in practice.

\h Because of the mathematical difficulties met when looking for exact solutions, often the analytic description must be obtained in an approximate way, and several other times one
must have recourse to a numerical solution of the (differential or integral) propagation equations both in their exact or approximate forms.

\h A very common approximation is the paraxial one\cite{Fresnel}, rather useful for obtaining analytic or numerical solutions. For example, it is by such an approximation that one
obtains the well-known Fresnel diffraction integral\cite{Fresnel}, which yields accurate (analytic and numerical) results for a large part of the proximal field region, as well as in
the transition region towards the distant field.

\h An important analytic solution forwarded by the Fresnel diffraction integral is the gaussian beam, while another one is the Bessel-Gauss beam, found by Gori et al.\cite{Gori} in
1987. The latter, endowed with a transverse profile in which the Bessel function is modulated by a gaussian function, can be regarded as an experimentally realizable version of the Bessel beam; indeed, the Bessel
beam is a quite noticeable {\em exact, non-diffracting solution} of the wave equation, but is associated with an infinite power flux (through any plane orthogonal to the propagation
axis), as it also happens, for instance, with plane waves.

\h Notwithstanding the fact that some analytic solutions do exist for the Fresnel diffraction integral, they are rare, and normally it is necessary to have recourse to numerical
simulations. This is particularly true when the mentioned integral is adopted for the description of beams generated by finite apertures, that is, of beams truncated in space.

\h The past attempts at an analytic description of truncated beams were based on a Fresnel integral: probably the best known of
them being the Wen and Breazele
method\cite{WenBreazele}, using superpositions of gaussian beams (with different waist sizes and positions) in order to
describe {\em axially symmetrical} beams truncated by circular
apertures. In that approach, those authors had to adopt a computational optimization process to get the superposition coefficients, and the beam waists and spot positions of the
various gaussian beams; actually, the necessity of a computational optimization to find out which beam superposition be adequate to describe a certain truncated beam is due to the
simple fact that the gaussian beams do not constitute an orthogonal basis...

\h To cope with the difficulty related with such a computational optimization, Ding and Zhang\cite{DingZhang} modified the method
by choosing since the beginning the beam waist values,
and then writing down a set of linear equations in terms of the gaussian beam superposition coefficients. However, in that new approach the nonhomogeneous terms are given by integrals
that, once more, cannot in general be easily evaluated in closed form.

\h In this paper we are going to show that an {\em analytic} description of important truncated beams can be obtained by means of Bessel-Gauss beam superpositions, whose coefficients
are got in a simple and direct way, without any need of numerical optimizations or of equation system solutions.

\h Indeed, our method is capable of yielding analytic solutions for the 3D evolution of Bessel beams, plane waves, gaussian beams, Bessel-Gauss beams, even when truncated by finite
apertures in the Fresnel regime.

\

\section{The Fresnel diffraction integral and some solutions}

\h In this paper, for simplicity's sake, we shall leave understood in all solutions the harmonic time-dependence term \ $\exp (-i\om t)$.

\h In the \emph{paraxial approximation}, an axially symmetric monochromatic wave field can be evaluated, knowing its shape on the $z=0$ plane, through the Fresnel diffraction integral
in cylindrical coordinates:

\bb \Psi(\rho,z) \ug \frac{-i k}{z}\, \exp  \left[ i \left( k z + \frac{k\rho^2}{2z} \right) \right]\,\int_{0}^{\infi}\,\Psi(\rho',0)\,\exp  \left( ik\frac{\rho'^2}{2z} \right)
\,J_0\left(k\frac{\rho\rho'}{z}\right)\rho' d\rho' \ , \label{fresnel} \ee

where $k=2\pi / \lambda$ is the wavenumber, and $\lambda$ the wavelength. In this equation, $\rho'$ recalls us that the integration is being performed on the plane $z=0$; thus,
$\Psi(\rho',0)$ does simply indicates the field value on $z=0$.

\h Let us consider a gaussian behaviour on $z=0$, that is to say, let us choose the ``exitation"

\bb \Psi(\rho',0) \ug \Psi_G(\rho',0) \ug  A\exp (-q\rho'^2) \; , \label{gz0} \ee

contained in ref.\cite{Fresnel}; with $A$ and $q$ constants {\em that can have a complex value}. For ${\rm Re}(q) \geq 0$, one gets the pretty known gaussian beam solution:

\bb \Psi_G(\rho,z) \ug -\frac{i k A}{2 z Q}\,\exp \left[i k\left(z + \frac{\rho^2}{2z} \right) \right] \exp \left[-\frac{k^2\rho^2}{4Qz^2}\right] \ , \label{gauss} \ee

where

\bb Q = q - ik/2z \label{Q} \; . \ee

\h Another important solution is obtained by considering on the $z=0$ plane the excitation given by

\bb \Psi(\rho',0) \ug \Psi_{BG}(\rho',0) \ug A J_0(\kr \rho')\exp (-q\rho'^{\,2}) \ , \label{bgz0} \ee

which, according to eq.(\ref{fresnel}), produces the so-called Bessel-Gauss beam\cite{Gori}:

\bb \Psi_{BG}(\rho,z) \ug -\frac{i k A}{2 z Q}\,\exp \left[i k\left(z + \frac{\rho^2}{2z} \right) \right] \, J_0\left(\frac{i k \kr \rho}{2zQ}\right) \exp
\left[-\frac{1}{4Q}\left(\kr^2 + \frac{k^2\rho^2}{z^2}\right) \right] \ , \label{bg} \ee

quantity $Q$ being given by eq.(\ref{Q}), and $\kr$ being a constant.\footnote{Quantity $\kr$ is the transverse wavenumber associated with a Bessel beam transversally modulated by
the gaussian function.}

 \h The Bessel-Gauss beam given by eq.(\ref{bg}) is particularly interesting since it can well be regarded as a realistic
version (experimentally speaking) of the ideal Bessel beam:

\bb \Psi_{B}(\rho , z) \ug  A J_0\left(\kr \rho \right)\exp  (ik_z z) \ , \label{bb} \ee

where $k_z = \sqrt{\om^2/c^2 - \kr^2}$.

\h The Bessel beam, eq.(\ref{bb}), is an exact solution to the wave equation and is known to possess the important characteristic of keeping its transverse behavior unchanged while
propagating, so to belong to the class of the nondiffracting beams. \ However, the ideal Bessel beam is endowed with an infinite power flux, and cannot be concretely generated.  By
contrast, the Bessel-Gauss beam, eq.(\ref{bg}), modulates in space the transverse behaviour of the Bessel beam by a gaussian function, getting a finite power flux.  The Bessel-Gauss
beam will no longer remain indefinitely undistorted, but nevertheless shows to possess a rather good resistance to diffraction\cite{Gori}.

\h The gaussian beam, eq.(\ref{gauss}), and the Bessel-Gauss, eq.(\ref{bg}), solutions are among the few solutions to the Fresnel
diffraction integral that can be got analytically. \
The situation gets much more complicated, however, when facing beams truncated in space by finite circular apertures:
For instance, a gaussian beam, or a Bessel beam, or a
Bessel-Gauss beam, truncated via an aperture with radius $R$. In this case, the upper limit of the integral in
eq.(\ref{fresnel}) becomes the aperture radius, and the analytic
integration becomes very difficult, requiring recourse to lengthy numerical calculations.

\h As we already mentioned, in their alternative attempt at describing truncated beams, Wen and Breazele\cite{WenBreazele} adopted superpositions of gaussian beams. To be, now, more
specific, those authors wrote down the solution for a wave equation in the paraxial approximation as:

\bb \Psi(\rho,z) \ug - \frac{i k}{2 z}\,\exp \left[i k\left(z + \frac{\rho^2}{2z} \right) \right]\, \dis{\sum_{n=1}^{N}}\, \frac{A_n}{Q_n}\,\exp \left[-\frac{k^2\rho^2}{4 Q_n
z^2}\right] \ , \label{sg} \ee

with

\bb Q_n = q_n - \frac{ik}{2z} \label{Qn} \; . \ee

\h Solution (\ref{sg}) is a superposition of $N$ gaussian beams. Its coefficients $A_ n$ and $q_n$ are to be obtained starting from the field existing on the $z=0$
plane, that is, starting from the initial excitation that we shall call $V(\rho)$. One therefore looks for $\Psi(\rho,0) = V(\rho)$.  From eq.(\ref{sg}) one gets

\bb V(\rho) \ug  \dis{\sum_{n=1}^{N}}\, A_n \exp \left( -q_n \rho^2  \right) \ . \label{sgz0} \ee

The initial field $V(\rho)$ can represent a beam (e.g., a plane wave, a gaussian beam,...) {\em truncated} by a circular aperture with radius $R$. \ To get the coefficients $A_ n$ and
$q_n$ from eq.(\ref{sgz0}), those authors had recourse to a {\em computational optimization} process in order to minimize their mean square error: Namely, to minimize the difference
between the desired function  $V(\rho)$ and the gaussian series in the r.h.s. of eq.(\ref{sgz0}). \ Such a method yields good results, provided that the exitation function $V(\rho)$
does not oscillate too much. But the coefficients $A_ n$ and $q_n$ are obtained in a strictly not algebraic manner, depending on the contrary on numerical calculations.

\h Let us be more specific also about the modification of Wen and Breazele's method introduced by Ding and Zhang\cite{DingZhang}.  They postulated the values of the parameters
$q_n$ and, by minimizing the mean square error between the desired function $V(\rho)$ and the gaussian series in eq.(\ref{sgz0}), arrived at a system of linear equations containing the
unknowns $A_n$ (the coefficients of the gaussian beam superposition), without needing a numerical optimization process. \ However, in the system of equations needed to determine the
coefficients $A_n$, the non-homogeneous terms consist in integrals that, depending on the field one wishes to truncate (i.e., depending on $V(\rho)$) can be difficult to be calculated analytically...

\h In the next Section we are going to propose a method, for the description of truncated beams, that appears to be noticeable for his simplicity and, in most cases, for its total
analiticity. Our method is based on Bessel-Gauss beam superpositions, whose coefficients can be directly evaluated without any need of computational optimizations or of solving any coupled equation systems.

\

\section{The method}

\h Let us start with the Bessel-Gauss beam solution, eq.(\ref{bg}), and consider the solution given by the following superposition of such beams:

\bb \Psi(\rho,z) \ug - \frac{i k}{2 \, z}\,\,\exp \left[i k\left(z + \frac{\rho^2}{2 \, z} \right) \right]\, \dis{\sum_{n=-N}^{N}}\, \frac{A_n}{Q_n}\, J_0\left(\frac{i \, k \,\kr
\,\rho}{2 \, z \, Q_n} \right)
    \, \exp \left[-\frac{1}{4 Q_n}\left(\kr^2 + \frac{k^2 \rho^2}{z^2}\right) \right] \; , \label{geral} \ee

quantities $A_n$ being constants, and $Q_n$ being given by eq.(\ref{Qn}), so that $Q_n = q_n - ik/2z $, where $q_n$ are constants that {\em can have complex values.}  Notice that in
this superposition all beams possess the same value of $\kr$.

\h Let us recall, incidentally, that all the beams we are considering in this work are important particular cases of the so-called Localized Waves (see \cite{Livro,AIEP,IEEE} and refs. therein; see also \cite{MRH2,CK}).

\h Our purpose is that solution (\ref{geral}) be able to represent beams truncated by circular apertures: As
announced, we are particularly interested in the analytic description of truncated beams of Bessel, Bessel-Gauss, gaussian and plane wave types.

\h Given one of such beams truncated at$z=0$ by an aperture with radius $R$, we have to determine the coefficients $A_n$ and $q_n$ in such a way that eq.(\ref{geral}) represents with
fidelity the resulting beam. \ If the truncated beam on the $z=0$ plane is given by $V(\rho)$, we have to obtain $\Psi(\rho,0) = V(\rho)$; that is to say:

\bb V(\rho) \ug   J_0(\kr \rho)\,\dis{\sum_{n=-N}^{N}} A_n e^{-q_n\rho^2} \; . \label{geralz0} \ee

The r.h.s. of this equation is nothing but a superposition of Bessel-Gauss beams, all with the same value $\kr$, at $z=0$ [namely, each one of such beams is written at $z=0$ according
to eq.(\ref{bgz0})].

\h Equation (\ref{geralz0}) will provide us with the values of the $A_n$ and $q_n$, as well as of $N$.  Once these values have been obtained, the field emanated by the finite circular
aperture located at $z=0$ will be given by eq.(\ref{geral}). \ Remembering that the $q_n$ can be complex, let us make the following choices:

\bb q_n \ug q_R + i q_{In} \label{qvalor} \; , \ee

where $q_R>0$ is the real part of $q_n$, having the {\em same value} for every $n$, and $q_{In}$ is the imaginary part of $q_n$ given by:

\bb q_{In} \ug - \frac{2 \pi}{L}\, n \; , \label{qi} \ee

where $L$ is a constant with the dimensions of a square length.

With such choices, and assuming $N \rightarrow \infi$, equation (\ref{geralz0}) gets written as

\bb V(\rho) \ug   J_0(\kr \rho)\,\exp \left(-q_{R} \rho^2 \right)\dis{\sum_{n=-\infi}^{\infi}} A_n \, \exp \left(i\frac{2 \pi n}{L}\rho^2\right) \; , \label{geralz02} \ee

which has then to be exploited for obtaining the values of $A_n$, \ $\kr$, \ $q_R$ and $L$.


\h Let us recall that aim of our method is describing some important truncated beams starting from the value of their near fields (i.e., of their fields in the Fresnel region).

\h In the cases of a truncated Bessel beam (TB) or of a truncated Bessel-Gauss beam (TBG), it results natural to choose quantity $\kr$ in eq.(\ref{geralz02}) to be equal to the
corresponding beam transverse wavenumber.

\h In the case of a truncated gaussian beam (TG) or of a truncated plane wave (TP), by contrast, it is natural to choose $\kr = 0 $ in eq.(\ref{geralz02}).

\h In all cases, the product

\bb \exp \left(-q_{R} \rho^2 \right)\dis{\sum_{n=-\infi}^{\infi}} A_n \, \exp \left(i\frac{2 \pi n}{L}\rho^2\right) \; ,
\label{expr} \ee

in eq.(\ref{geralz02}) must represent:

(i) a function $\circrm (\rho/R)$, in the TB or TP cases;

(ii) a function $\exp \left(-q \, \rho^2 \right)\,\circrm (\rho/R)$, that is, a $\circrm$ function multiplied by a gaussian function, in the TBG or TG cases. Of course (i) is a
particular case of (ii) with $q=0$. \ It may be useful to recall that the $\circrm $-function is the step-function in the cylindrically symmetrical case. Quantity $R$ is still the aperture radius, and $\circrm (\rho/R)
= 1$ when $0 \leq \rho \leq R$, and equals $0$ in the contrary case.

\h Let us now show how expression (\ref{expr}) can approximately represent the above functions, given in (i) and (ii).

\h To such an aim, let us consider a function $G(r)$ defined on an interval $|r| \leq L/2$ and possessing the Fourier expansion:

\bb G(r) \ug \sum_{n=-\infi}^{\infi} A_n \exp (i 2 \pi n r / L) \;\;\;  {\rm for} \;\;\;  |r| \leq L/2  \label{G} \ee

where $r$ and $L$, having the dimensions of a square length, will be expressed in square meters ($m^2$).

\h Suppose now the function $G(r)$ to be given by

\bb
 G(r) \ug \left\{\begin{array}{clr}
& \exp \left(q_{R} \, r \right)\,\exp \left(-q \, r \right) \;\;\; {\rm for} \;\;\; |r| \leq R^2  \\

\\

&0 \;\;\; {\rm for}\;\;\; R^2 < |r| < L/2  \ ,
\end{array} \right.  \label{G1}
 \ee

where $q$ is a given constant. 


\h In this case, the coefficients $A_n$ in the Fourier expansion of de $G(r)$ will be given by:

\bb
\begin{array}{clr}

A_n \ug & \dis{\frac{1}{L} \, \int_{-R^2}^{R^2} \, \exp \left(q_R r \right)\,\exp \left(-q \, r \right)\, \exp \left(- i \frac{2\pi n}{L} r
\right)\,dr} \\

\\

\ug & \dis{\frac{1}{L \, (q_R-q) - i 2\pi n}} \left\{\exp \left[\left(q_R-q - i\frac{2\pi}{L} n \right)R^2 \right] - \exp \left[ - \left(q_R-q - i\frac{2\pi}{L} n \right)R^2
\right]\right\} \ ,  \end{array} \label{An1} \ee





\h Writing now

\bb r = \rho^2  \label{subs} \; , \ee

in eqs.(\ref{G},\ref{G1}) we shall have that

\bb
 \sum_{n=-\infi}^{\infi} A_n \exp (i 2 \pi n \rho^2 / L) \ug \left\{\begin{array}{clr}
& \exp \left(q_{R} \, \rho^2 \right)\,\exp \left(-q \, \rho^2 \right) \;\;\; {\rm for}\;\;\; |\rho| \leq R  \\

\\

&0 \;\;\; {\rm for}\;\;\; R < |\rho| < \sqrt{L/2} \ ,
\end{array} \right.  \label{G3}
 \ee

where the coefficients $A_n$ are still given by eq.(\ref{An1}). \ One recognizes that the l.h.s. of eq.(\ref{G3}) is the term multiplying the gaussian in
expressio (\ref{expr}).

\h The l.h.s. of  eq.(\ref{G3}), which depends on $\rho^2$, will not be exactly periodical: But it will re-assume the values, assumed in the fundamental interval ($|\rho|<\sqrt(L/2)$),
in shorter and shorter further intervals, with decreasing spatial ``periodicity", for $|\rho|>\sqrt(L/2)$. \ In such a way, with $\rho \geq 0$, expression (\ref{expr}) becomes:




\bb
 \exp \left(-q_{R} \rho^2 \right)\,\sum_{n=-\infi}^{\infi} A_n \exp (i 2 \pi n \rho^2 / L) \ug \left\{\begin{array}{clr}
& \exp \left(-q \, \rho^2 \right) \;\;\; {\rm for}\;\;\; 0 \leq \rho \leq R  \\

\\

& 0 \;\;\; {\rm for}\;\;\; R < \rho \leq \sqrt{L/2} \\

\\

& \exp \left(-q_{R} \rho^2 \right)\,f(\rho) \approx 0 \;\;\; {\rm for}\;\;\; \rho > \sqrt{L/2}  \ ,
\end{array} \right.  \label{16}
 \ee

where $f(\rho)$ is a function existing on decreasing space intervals, and assuming $\exp[(q_R-q)R^2]$ as its maximum values.  Since $\sqrt{L/2}>R$, for suitable choices of $q_R$ e $L$, we shall have that $\exp(-q_{R} \rho^2)\,f(\rho)\approx 0$ for $\rho \geq \sqrt{L/2}$.

\h Therefore, we  get that

\bb  \exp \left(-q_{R} \rho^2 \right)\,\sum_{n=-\infi}^{\infi} A_n \exp (i 2 \pi n \rho^2 / L) \, \approx \, \exp \left(-q \, \rho^2 \right)\,\circrm (\rho/R) \label{expr2} \;\; , \ee

which corresponds to the case (i), when $q=0$, and to the case (ii).

\h Let us recall once more that the $A_n$ are given by eqs.(\ref{An1}).

\h On the basis of what shown before, we have now in our hands a very efficient method for describing important beams, truncated by finite apertures: Namely, the TB, TG, TBG, and TP beams.\ Indeed, it is enough to choose the desired field, truncated by a
circular aperture with radius $R$, and describe it at $z=0$ by our eq.(\ref{geralz02}). Precisely:

\begin{itemize}
    \item{In the TBG case: the value of $\kr$ in eq.(\ref{geralz02}) is the transverse wavenumber of the Bessel beam modulated by the gaussian function; \ $A_n$ is given in eq.(\ref{An1}); \ $q$ is related to the gaussian function
        width at $z=0$. \ The value $L$ and $q_R$, and the number $N$ of terms in the series (\ref{geralz02}), are chosen so as to guarantee a faithful description of the beam at $z=0$ when truncated by a circular aperture with radius $R$.}
    \item{In the TB case: the procedure will be the same as for TBG, but with $q=0$}
    \item{In the TG case: the procedure is still the same as for TBG, but with $\kr = 0$}
    \item{In the TP case: the procedure is once more similar, but using this time $\kr = 0$ e $q=0$}
\end{itemize}

\h Finally, {\em once the chosen beam is described on the truncation plane ($z=0$), the beam emanated by the finite aperture  will be given by solution (\ref{geral})}.

\h It is important to notice at this point that, for a given truncation, innumerable sets of values of $q_R$ and $L$ exist which yield a faithful description of the truncated field. The choice is made in order that the solution given by the series (\ref{geralz02}) has good convergence properties.  Of course, one will use a finite number $2N+1$ of terms in the mentioned series, with $ -N \leq n \leq N $.

\h Let us go on to some {\em examples.}

\section{Applying the method}

\h In this Section we shall apply our method, as we already said, to situations in which important truncated beams appear. Notice
that below we shall always assume a wavelength of 632.8 nm.

\subsection{Analytic description of the truncated Bessel beam}

\h Let us start from a Bessel beam, truncated at $z=0$ by a circular aperture with radius $R$; that is to say,
from $\Psi_{TB}(\rho,0) \ug J_0(\kr \rho)\,\circrm (\rho/R)$.

\h Let us choose $R=3.5\;$mm, and the transverse wavenumber $\kr = 4.07\cdot 10^4 \; {\rm m}^{-1}$, which corresponds to a beam
spot with radius approximatively equal to $\Delta\rho = 59\;\mu$m (while $\lambda = 632.8 \;$nm, as always).

\h At $z=0$ the field is described by eq.(\ref{geralz02}), where the $A_n$ are given by eq.(\ref{An1}) and where $q=0$. In this case, a quite good result can be got by the choice $L =3R^2$, \ $q_R = 6/L$  and  $N=23$. Let us repeat that, since such a choice is not unique, very many alternative set of values $L$ and $q_R$ exist, which yield as well excellent results.

\h
ure 1 shows the field given by eq.(\ref{geralz02}): it represents with high fidelity the Bessel beam truncated at $z=0$.

\begin{figure}[!h]
\begin{center}
 \scalebox{.25}{\includegraphics{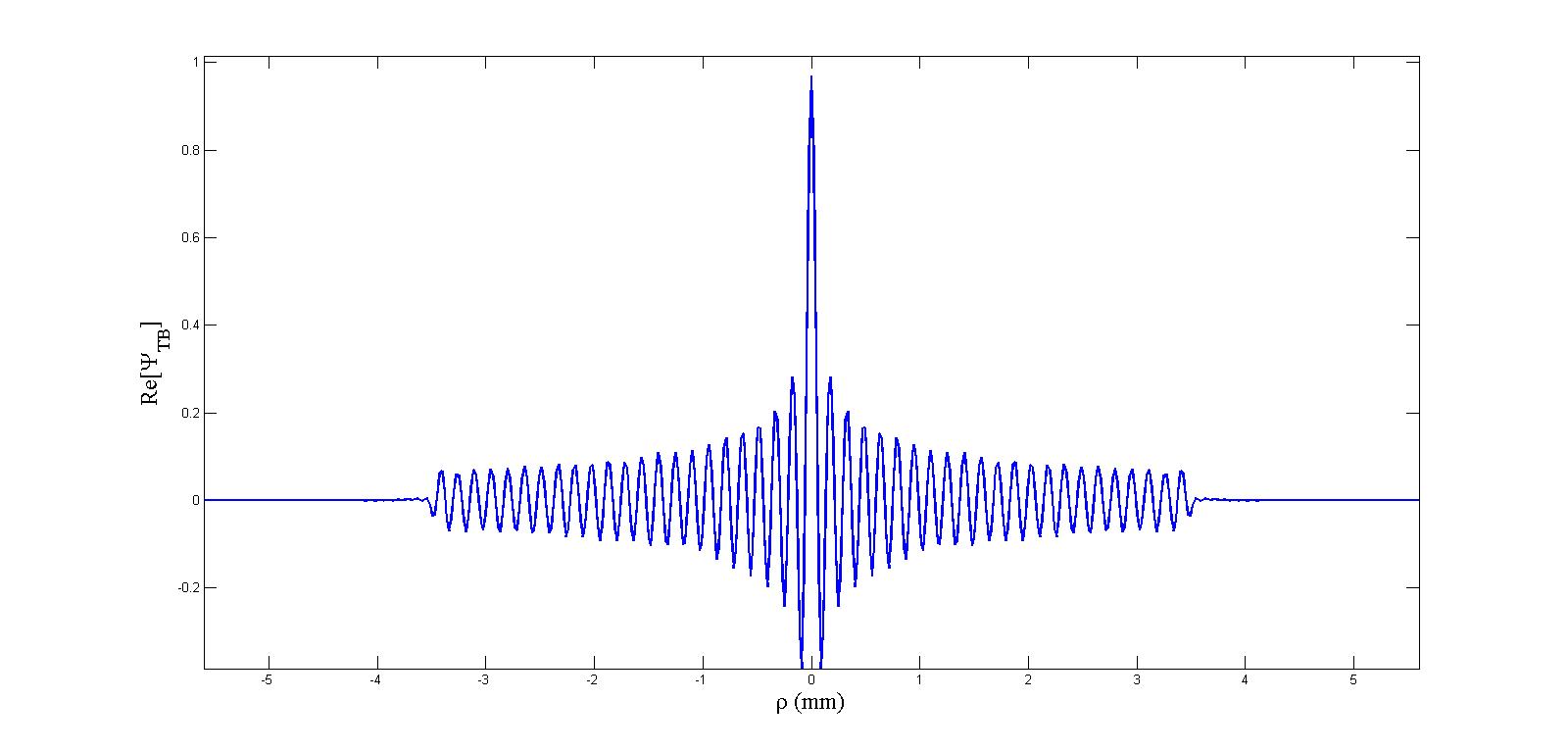}}
\end{center}
\caption{Field given by eq.(\ref{geralz02}), representing a Bessel beam at $z=0$, with $\kr = 4.07\cdot 10^4 \; {\rm m}^{-1}$ and truncated by a finite circular aperture with radius $R=3.5\;$mm.  The coefficients $A_n$ are given by eq.(\ref{An1}), with $q=0$, \ $L = 3R^2$, \ $q_R = 6/L$ and $N=23$.} \label{fig1}
\end{figure}

\h The resulting field, emanated by the aperture, is given by solution (\ref{geral}), and its intensity is shown in Fig.2. One can see that the result really corresponds to a Bessel beam truncated by a finite aperture. Figure 3 depicts the orthogonal projection of the same result.

\begin{figure}[!h]
\begin{center}
 \scalebox{.3}{\includegraphics{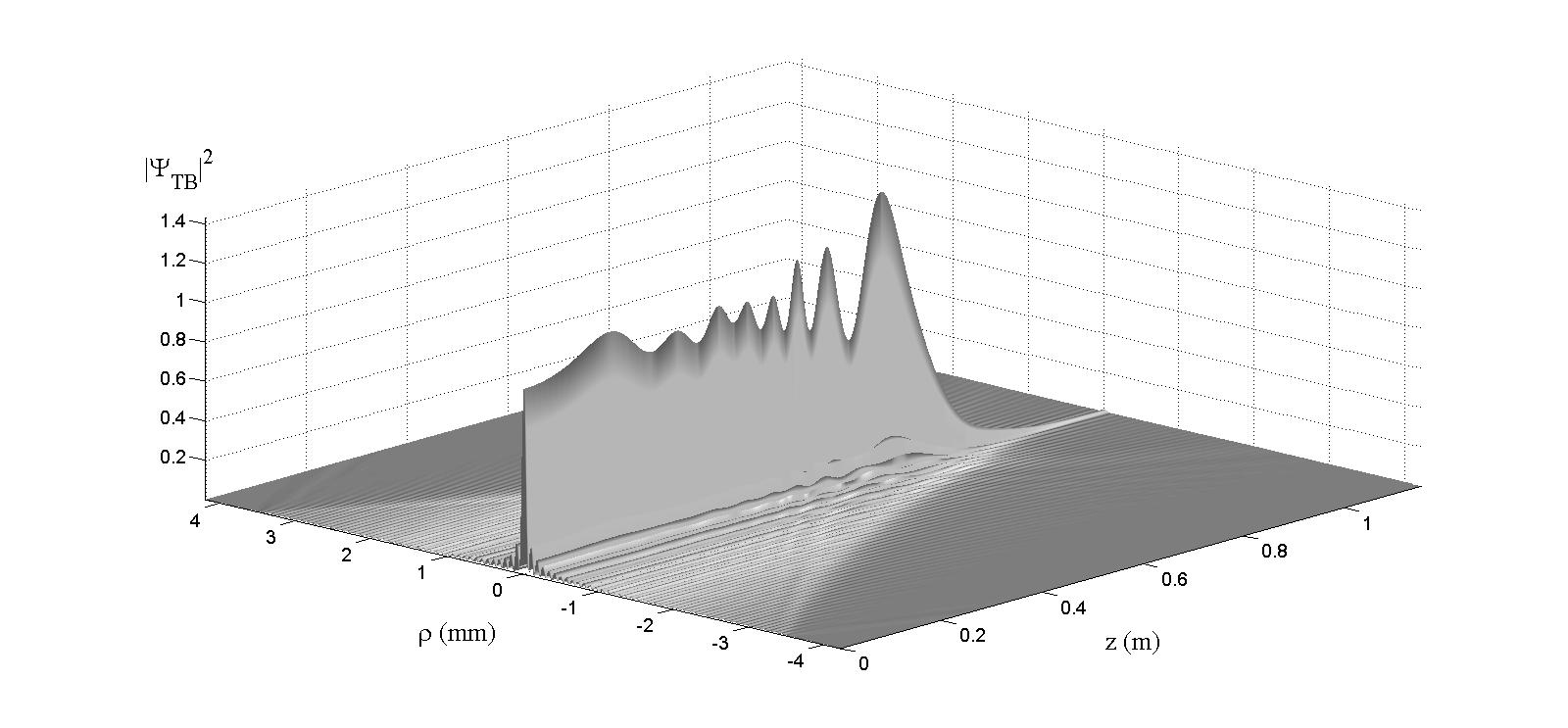}}
\end{center}
\caption{Intensity of a Bessel beam truncated by a finite aperture, as given by solution (\ref{geral}). } \label{fig2}
\end{figure}

\begin{figure}[!h]
\begin{center}
 \scalebox{.23}{\includegraphics{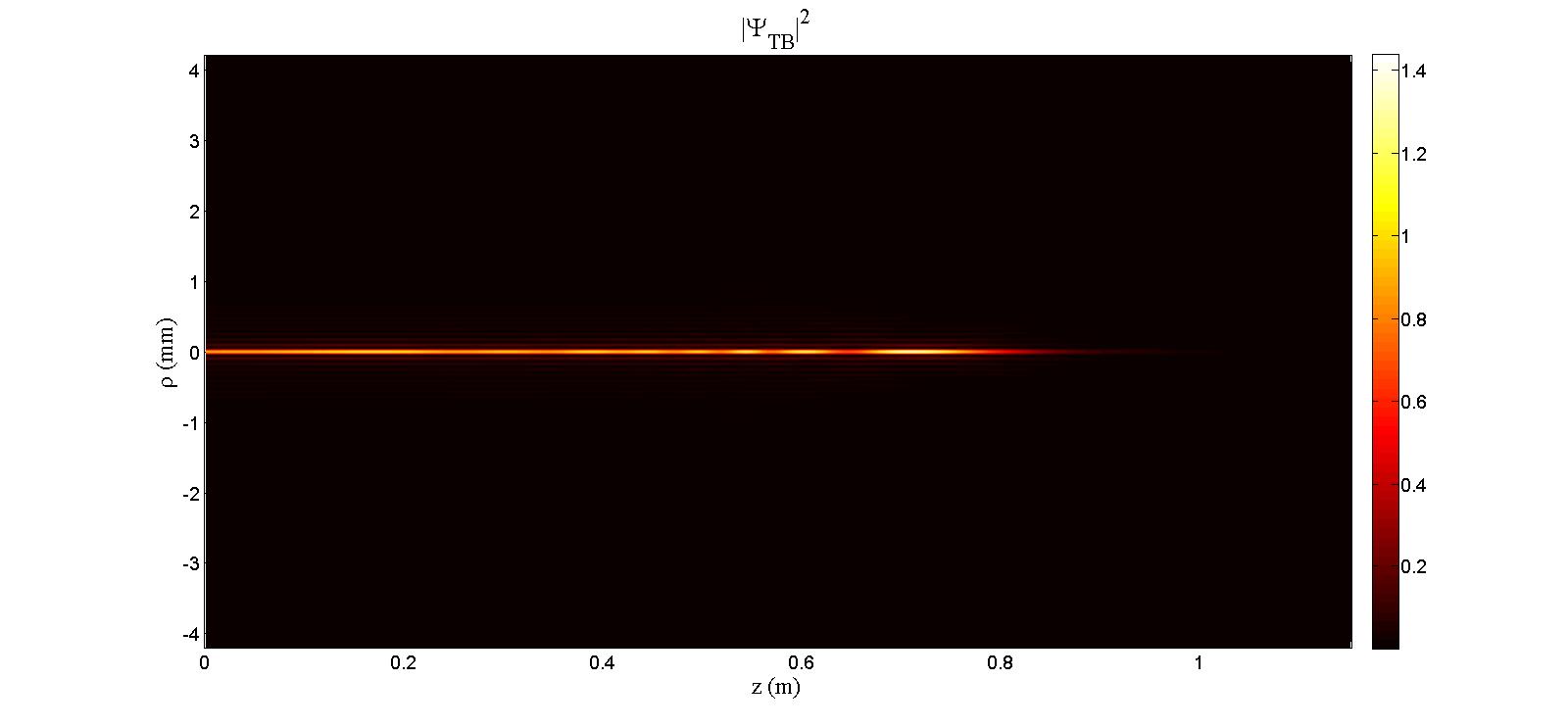}}
\end{center}
\caption{Orthogonal projection of the intensity shown in Fig.2.} \label{fig3}
\end{figure}

\h Increasing $N$, that is, increasing the number of terms in the series (\ref{geral}) which expresses the resulting field, while keeping the same values for $L$ and $q_R$,
the spatial shape of the obtained field practically will not change; but there will become more evident the rapid oscillations that occur at the beam crest, i.e., for $\Psi_{TB}(\rho=0,z)$. This is shown
in Fig.4, were we used $N=500$.

\begin{figure}[!h]
\begin{center}
 \scalebox{.23}{\includegraphics{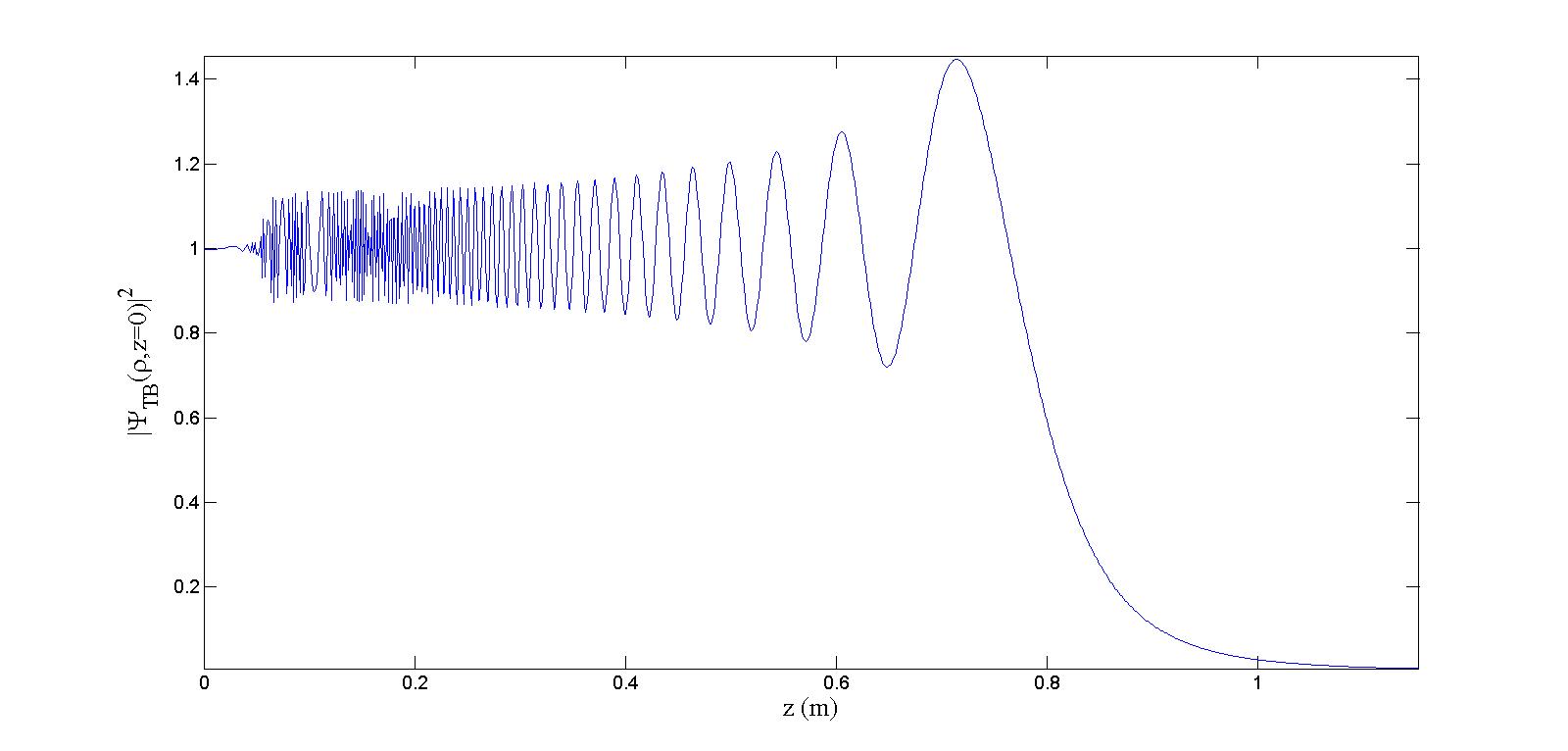}}
\end{center}
\caption{Oscillations of the field intensity on the $z$ axis, when adopting $N=500$ in eq.(\ref{geral}).} \label{fig4}
\end{figure}


\subsection{Analytic description of the truncated gaussian beam}

\h Let us go on now to consider a gaussian beam truncated at $z=0$; that is, $\Psi_{TG}(\rho,0) \ug \exp (-q\rho^{\,2})\,
\circrm (\rho/R)$, whose initial intensity spot radius is $\Delta\rho=59\;\mu$m, and therefore $q= 1/(2\Delta\rho^2) = 144
\cdot 10^{6}\; {\rm m}^{-1}$. The radius of the circular aperture is equal to the beam spot radius, i.e., $R=59\;\mu$m, \
while, as always, $\lambda = 632.8 \;$nm.

\h The situation at $z=0$ is still described by eq.(\ref{geralz02}), with $\kr=0$, where the $A_n$ are given by eq.(\ref{An1}). Now, a good result can be obtained, for instance, by using the values $L = 4R^2$, \ $q_R = 8/L$ and $N=81$.


\h In Fig.5 we show the field given by eq.(\ref{geralz02}), in the case of a gaussian beam truncated at $z=0$.  \ The dotted line depicts the ideal gaussian curve, without truncation.

\begin{figure}[!h]
\begin{center}
 \scalebox{.3}{\includegraphics{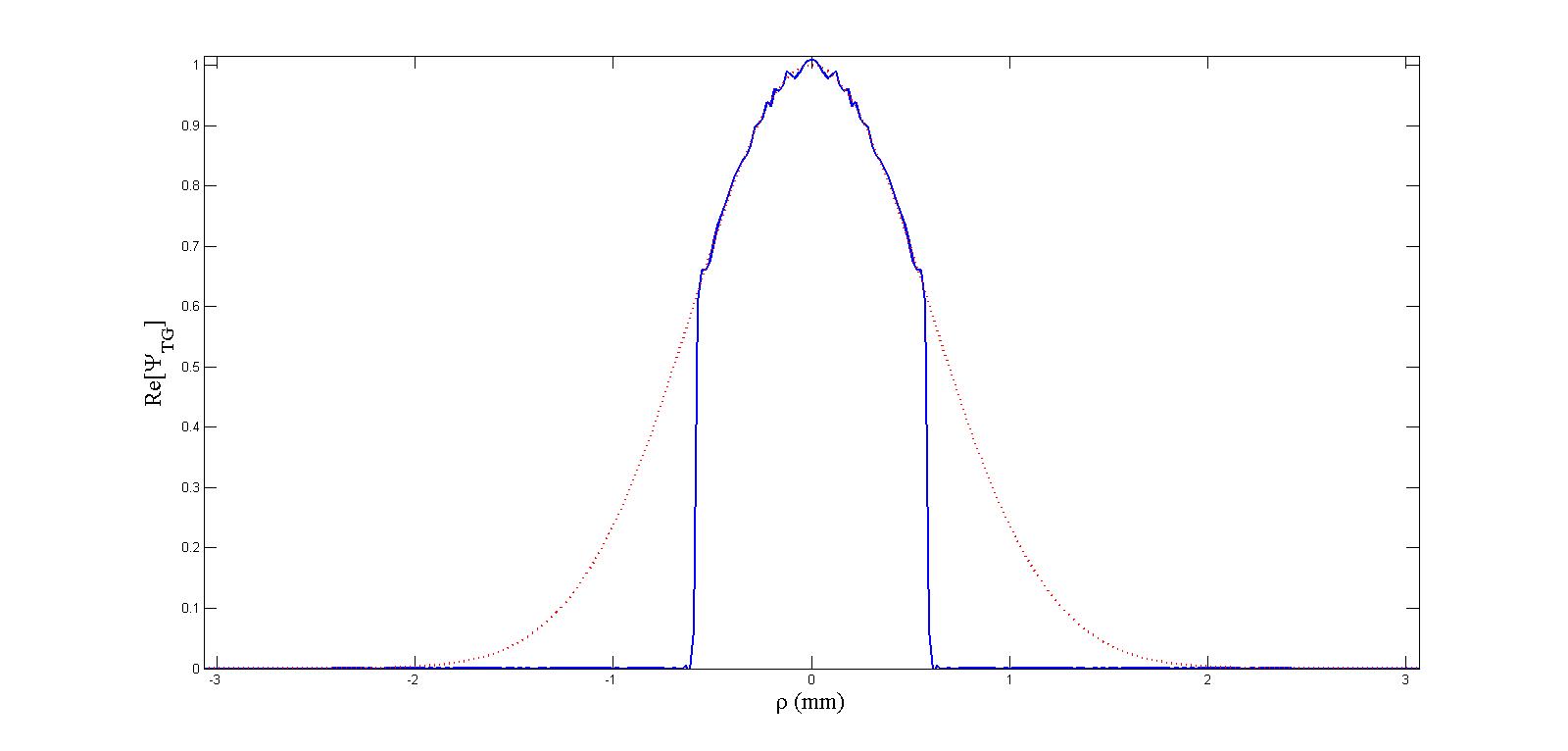}}
\end{center}
\caption{In this figure it is shown the field given by eq.(\ref{geralz02}), when representing a gaussian beam truncated at $z=0$.  \ The dotted line depicts the ideal gaussian curve, when truncation is absent.} \label{fig5}
\end{figure}

\h The resulting field, emanated by the finite aperture, is given by the solution (\ref{geral}), and Fig.6 shows its square
magnitude. In Fig.7 we show the corresponding orthogonal projection.

\begin{figure}[!h]
\begin{center}
 \scalebox{.3}{\includegraphics{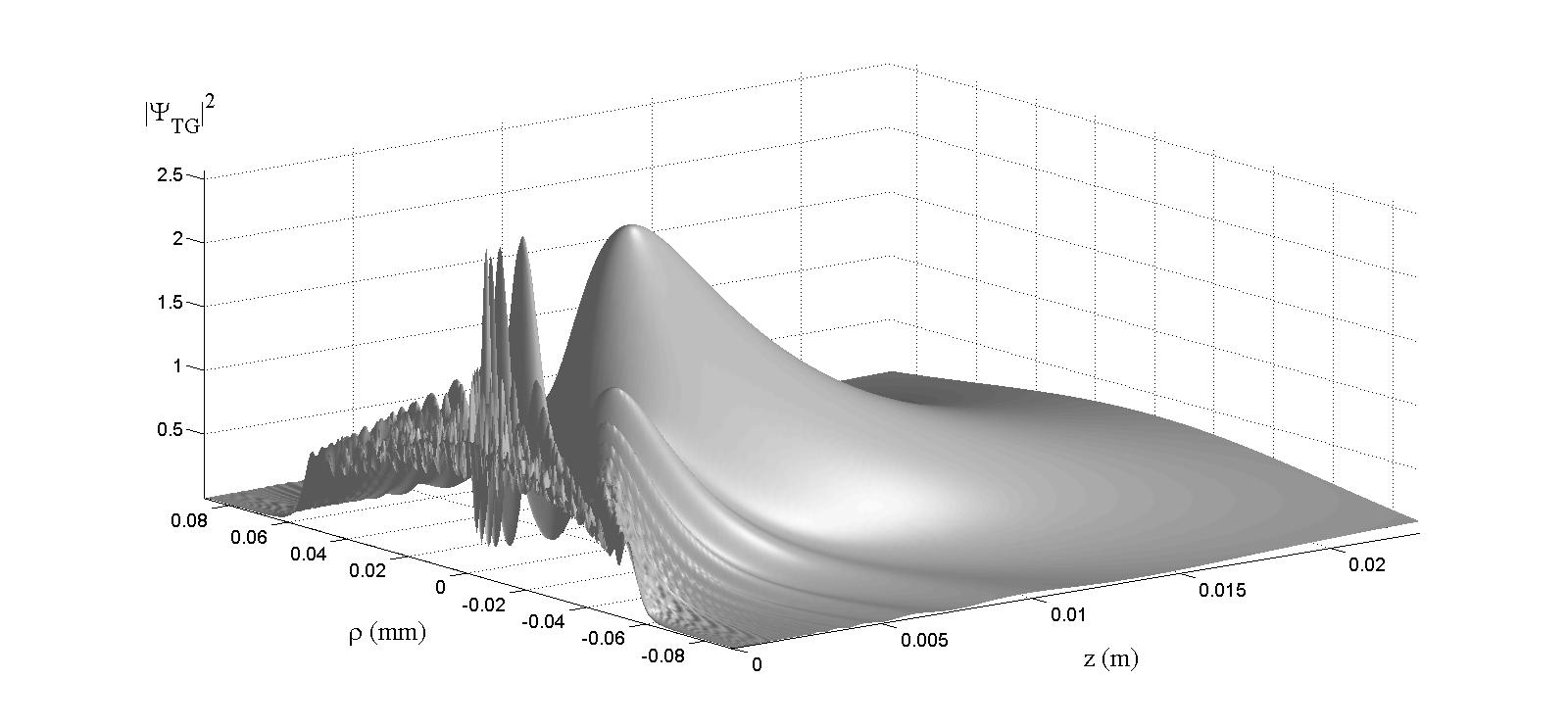}}
\end{center}
\caption{\h This Figure shows the square magnitude of the field emanated by a finite aperture, in the case of a gaussian beam, according to the solution (\ref{geral}).} \label{fig6}
\end{figure}

\begin{figure}[!h]
\begin{center}
 \scalebox{.6}{\includegraphics{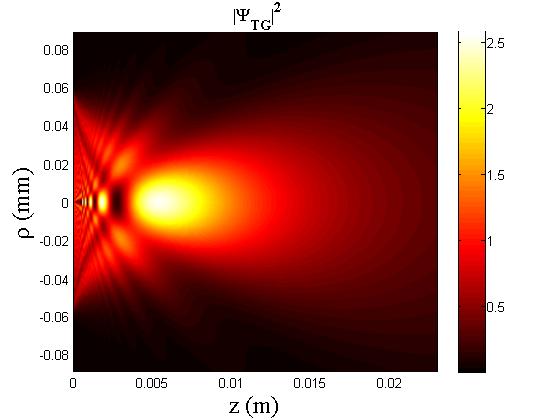}}
\end{center}
\caption{This Figure shows the orthogonal projection coresponding to the previous Figure.} \label{fig7}
\end{figure}

\newpage

\subsection{Analytic description of a truncated Bessel-Gauss beam}

\h Let us now consider the interesting case of a Bessel-Gauss beam truncated by a circular aperture of radius $R$; that is,  $\Psi_{TBG}(\rho,0) \ug J_0(\kr \rho)\exp (-q\rho^{\,2})\,\circrm
(\rho/R)$, \ with \ $\kr = 4.07\cdot 10^4 \; {\rm m}^{-1}$, \ $q=1.44 \cdot 10^6 {\rm m}^{-1}$, \  $R = 1\;$mm, \ and
$\lambda = 632.8 \;$nm.

\h The situation at $z=0$ is described by eq.(\ref{geralz02}), where the $A_n$ are given by relations (\ref{An1}). A very good result can be otained, e.g., by adopting the values $L =10\,R^2$, \ $q_R = q$ and $N=30$. \ Figure 8 shows the field in  eq.(\ref{geralz02}), in the present case of a Bessel-Gauss beam truncated at $z=0$.

\h The resulting field emanated by the finite aperture is given by solution (\ref{geral}), and Fig.9 shows its square magnitude. Figure 10 depicts the orthogonal projection for this case.

\begin{figure}[!h]
\begin{center}
 \scalebox{.3}{\includegraphics{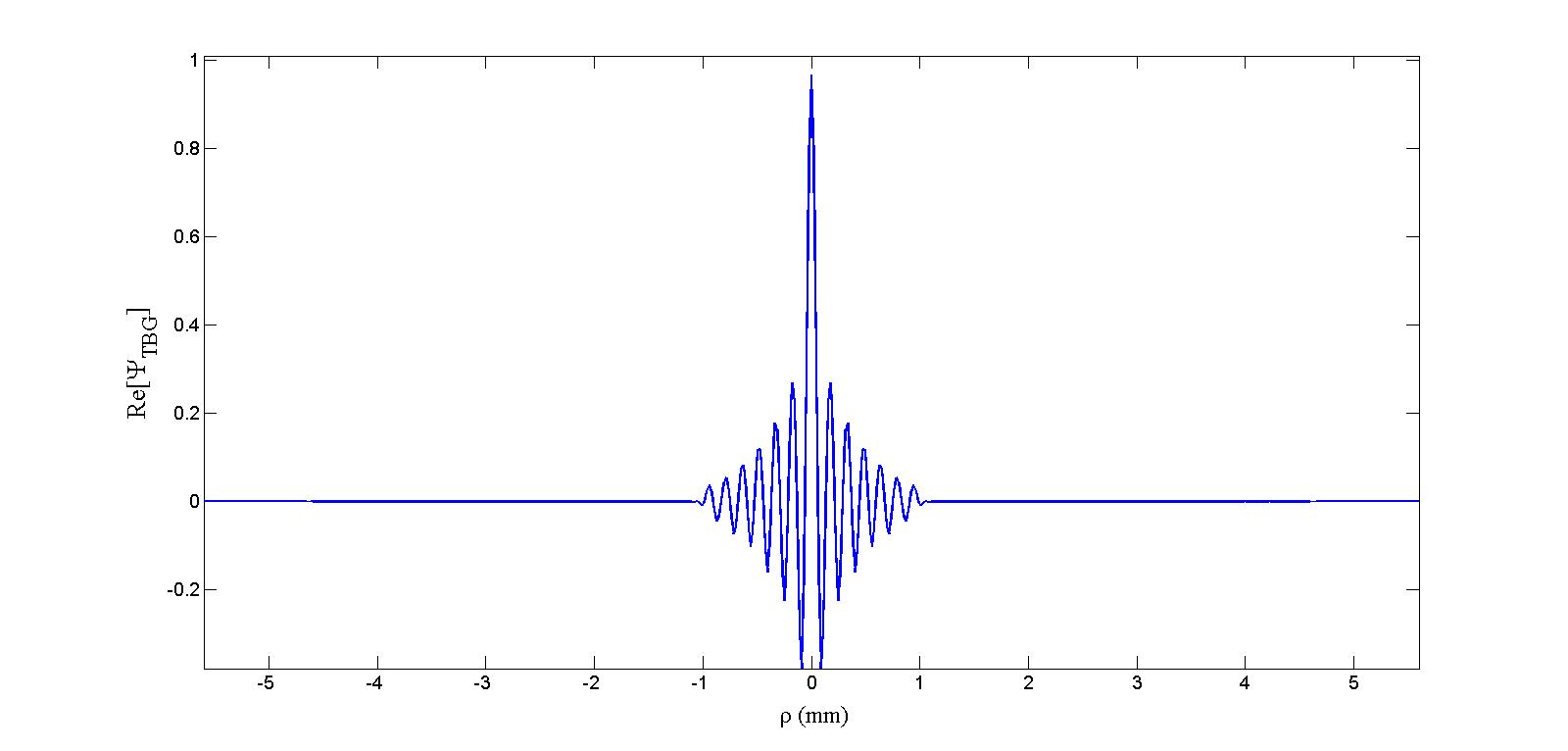}}
\end{center}
\caption{This Figure shows the field in eq.(\ref{geralz02}), in the case now of a {\em Bessel-Gauss beam} truncated at $z=0$. Here we adopted the values $L =10\,R^2$, \ $q_R = q$ and $N=30$. See the text.} \label{fig8}
\end{figure}

\begin{figure}[!h]
\begin{center}
 \scalebox{.3}{\includegraphics{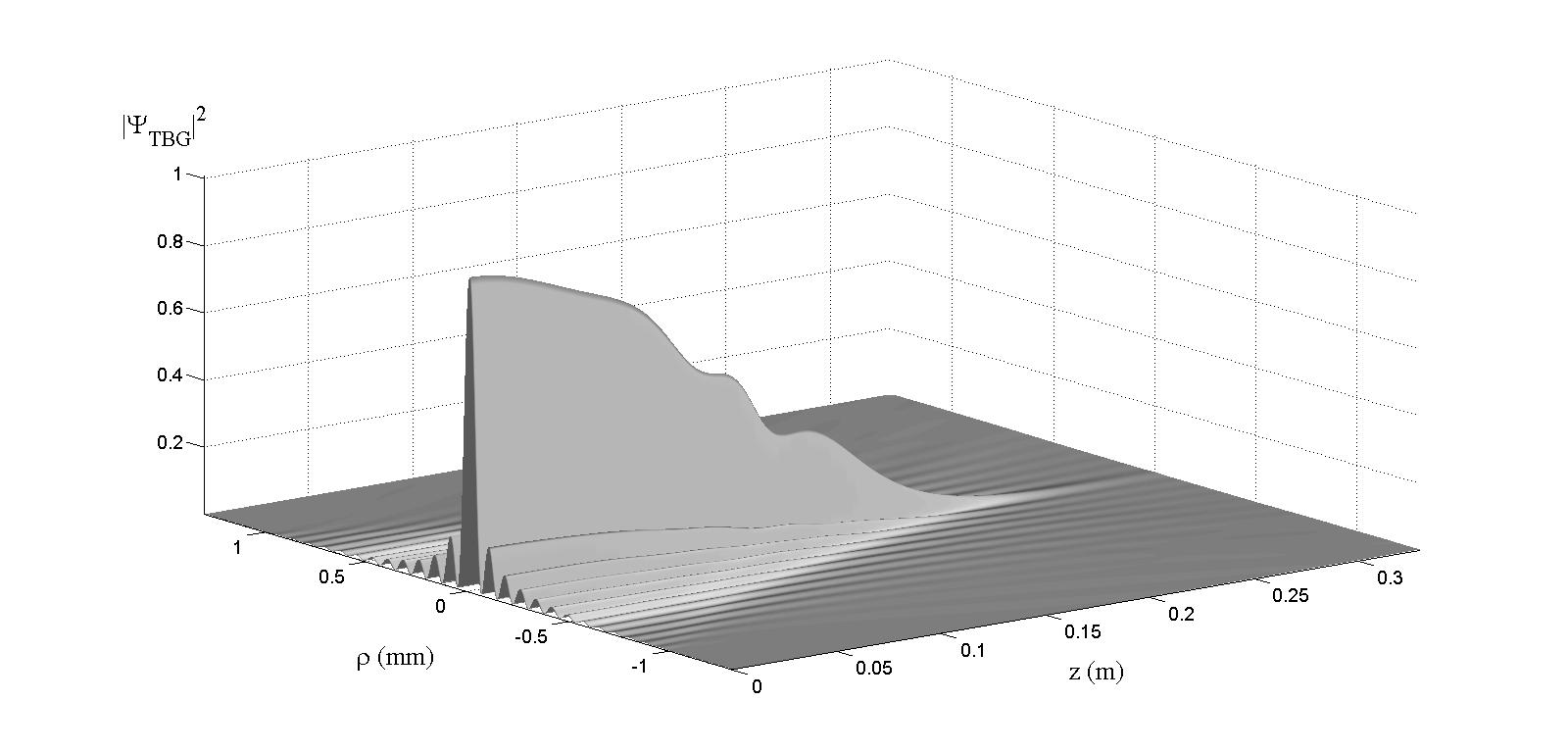}}
\end{center}
\caption{This Figure shows the square magnitude of the field emanated by a finite aperture in the case of a truncated Bessel-Gauss beam, represented by solution (\ref{geral}).} \label{fig9}
\end{figure}

\begin{figure}[!h]
\begin{center}
 \scalebox{.3}{\includegraphics{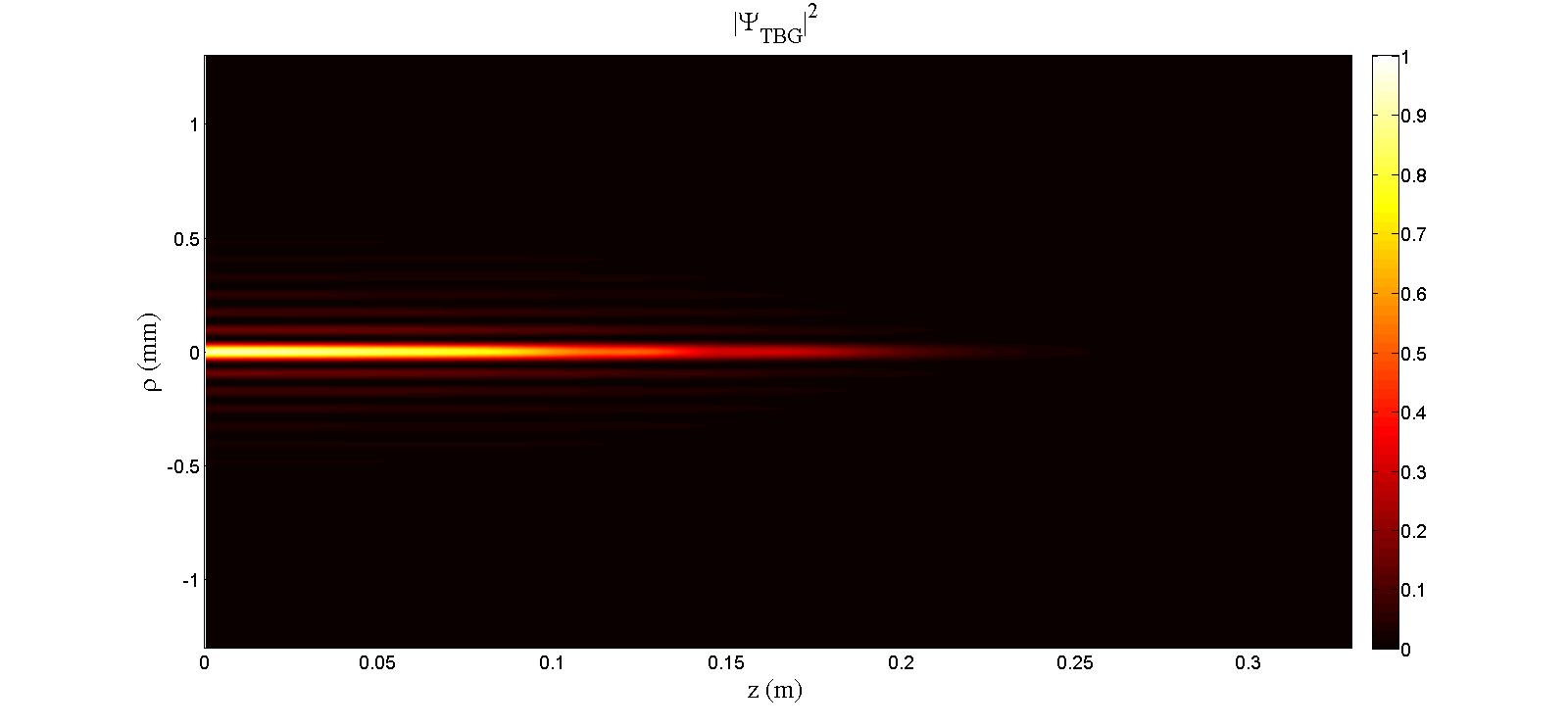}}
\end{center}
\caption{This Figure depicts the orthogonal projection for the case in the previous Figure.} \label{fig10}
\end{figure}

\

\

\newpage

\subsection{Analytic solution of a truncated plane wave}

\h Consider now the case of a plane wave truncated by a circular aperture em $z=0$, that is,
$\Psi_{TP}(\rho,0) \ug \circrm (\rho/R)$, were we choose $R=1\;$mm \ and $\lambda = 632.8 \;$nm.

\h Once more, eq.(\ref{geralz02}) describes the field at $z=0$, with $\kr=0$, the coefficients $A_n$ being given by  relations (\ref{An1}), with $q=0$. A good result can be got adopting, e.g., the values $L = 6\,R^2$, \ $q_R = 8/L$ and $N=150$.

\h Figure 11 shows the field in eq.(\ref{geralz02}), in the present case of a plane wave truncated at $z=0$.

\h The resulting field emanated by the finite aperture is given by solution (\ref{geral}), and its square magnitude is shown in Fig.12.  Figure 13 depicts the corresponding orthogonal projection.

\begin{figure}[!h]
\begin{center}
 \scalebox{.3}{\includegraphics{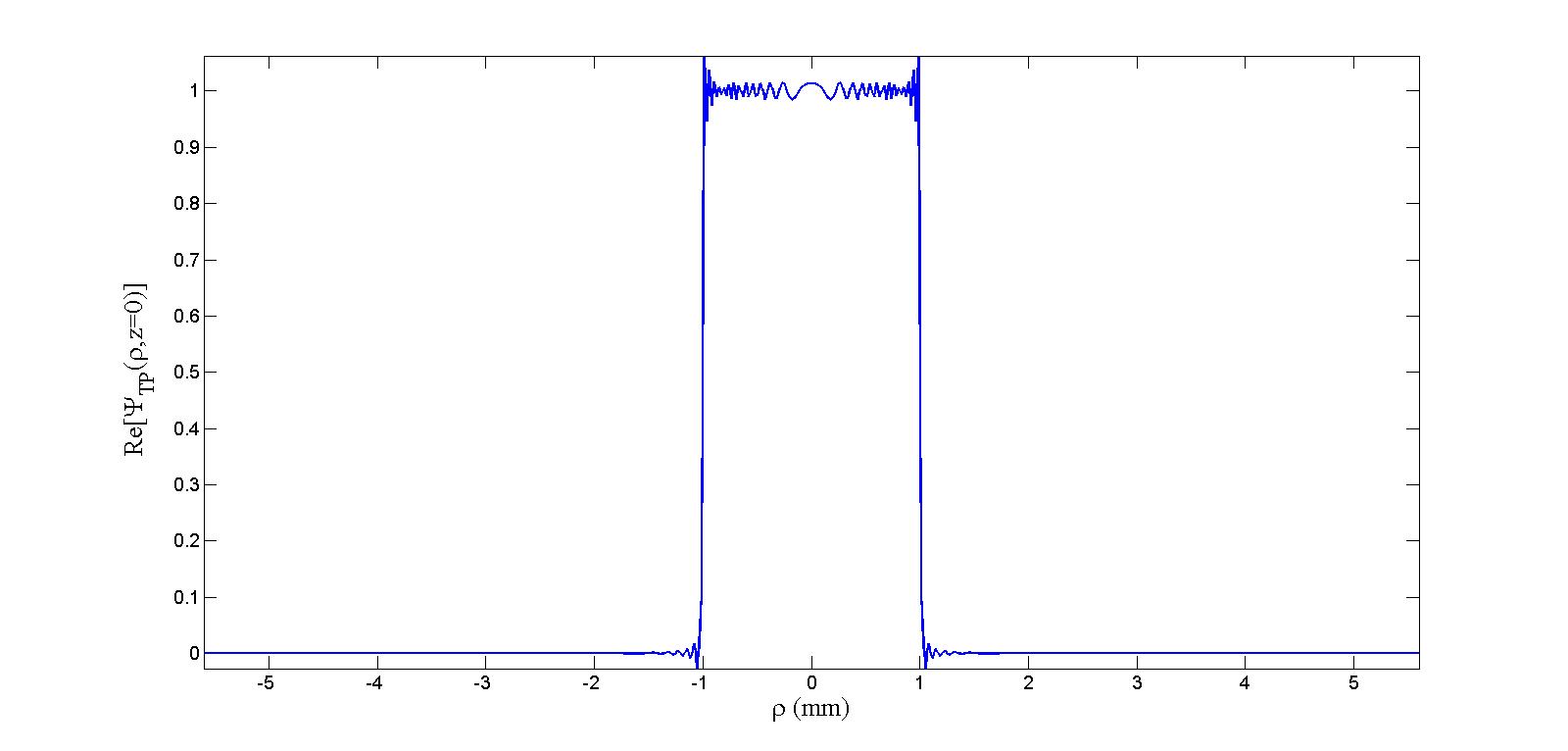}}
\end{center}
\caption{This Figure shows the field at $z=0$, with $\kr=0$, as given by eq.(\ref{geralz02} in the new case of a plane wave truncated at $z=0$. We have here adopted the values $L = 6\,R^2$, \ $q_R = 8/L$ and $N=150$. the coefficients $A_n$ being given by relations (\ref{An1}), with $q=0$.
} \label{fig11}
\end{figure}

\begin{figure}[!h]
\begin{center}
 \scalebox{.3}{\includegraphics{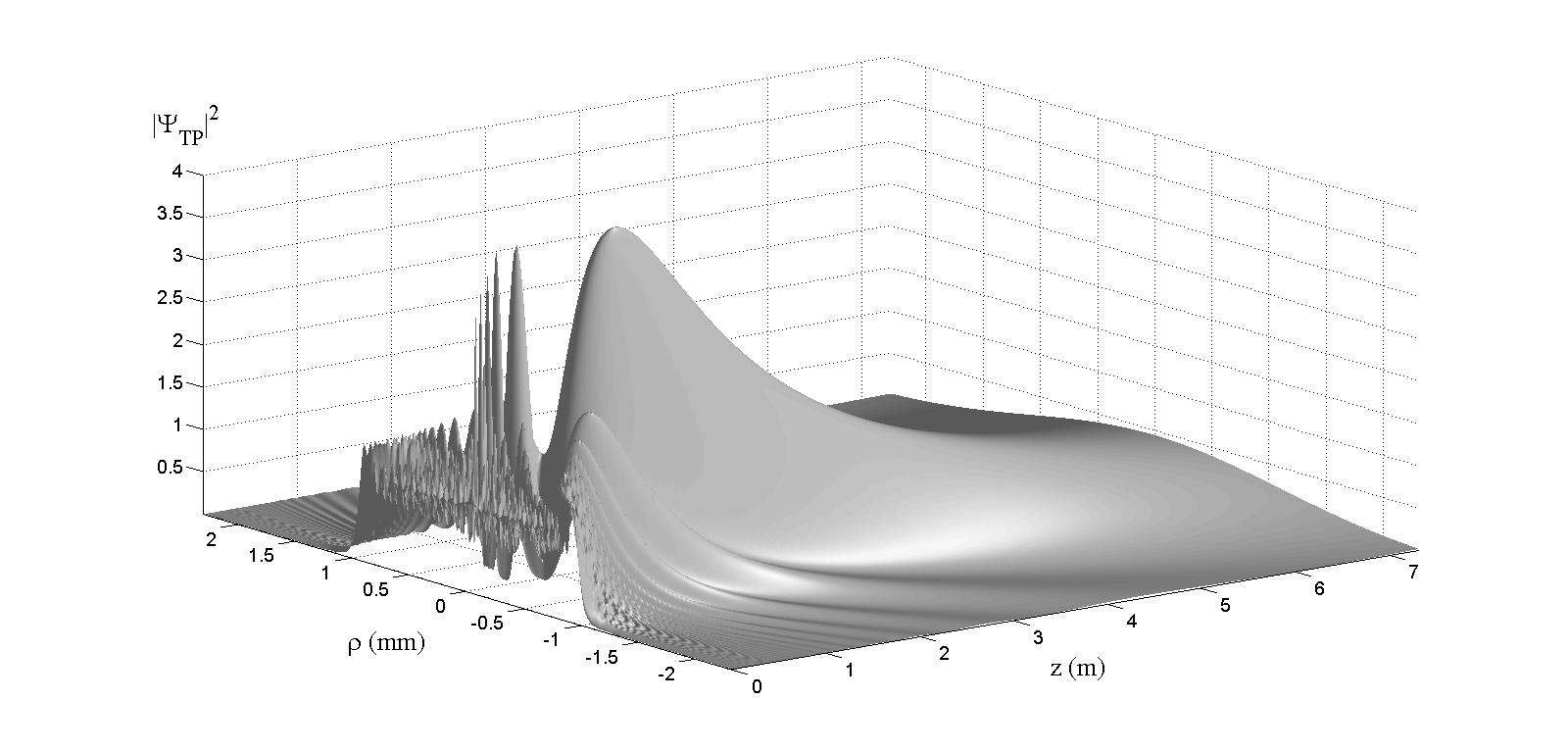}}
\end{center}
\caption{This Figure shows the square magnitude of the resulting field, emanated by the finite aperture, as given by solution (\ref{geral}) in the present case of a truncated plane wave.} \label{fig12}
\end{figure}

\begin{figure}[!h]
\begin{center}
 \scalebox{.3}{\includegraphics{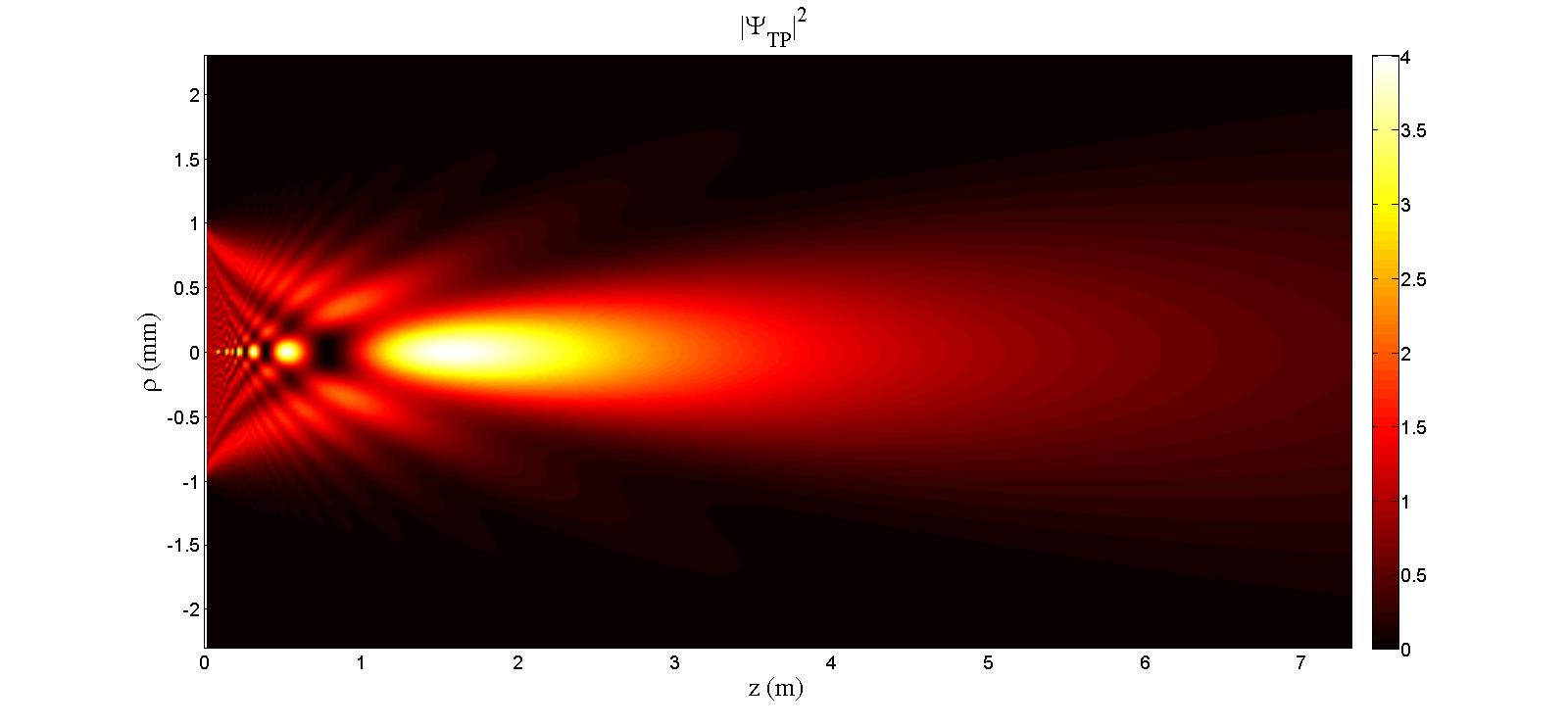}}
\end{center}
\caption{The orthogonal projection corresponding to the previous Figure.} \label{fig13}
\end{figure}

\section{Conclusions}

In this paper, starting from suitable superpositions of Bessel-Gauss beams\cite{Gori},
we have constructed a simple, effective method for the {analytic} description, in the Fresnel region, of
important beams truncated by finite apertures.

\h The solutions obtained by our method, and representing truncated Bessel beams, truncated gaussian beams, truncated
 Bessel-Gauss beams and truncated plane waves, fully agree with the known results obtained by lengthy
 numerical evaluations of the corresponding Fresnel diffraction integrals. (Incidentally, let us mention that
 all the beams considered in this work are important particular cases of the so-called
Localized Waves\cite{Livro,AIEP,IEEE,MRH2}).

\h At variance with the previous Wen and Breazele's approach\cite{WenBreazele} (which uses a computational
method of numerical optimization to obtain gaussian beam superpositions describing truncated beams), and even at
variance with Ding and Zhang's approach\cite{DingZhang} (which is an improved version of Ref.\cite{WenBreazele}),
our method does not need any numerical optimizations, nor the numerical solution of any coupled equation systems.

\h Indeed, the simpler method exploited in this paper is totally analytic, and directly applies to the beams
considered above, as well as to many other beams that are being investigated and will be presented elsewhere:
like truncated higher order Bessel and Bessel-Gauss beams; or beams truncated by circular apertures; or beams
truncated and modulated by convergent/divergent lenses; etc.  In particular we have applied this method to
remote sensing by microwaves (cf.,e.g., Ref.\cite{arXiv}), constructing finite antennas which emit truncated Bessel
beams with the required characteristics (patent pending). Of course, this method works in Electromagnetism (Optics,
Microwaves,...), as well as in Acoustics.

\h Let us stress that one of the main byproducts of our mathematical method is that by it one can get
in few seconds, or minutes, high-precision results which could otherwise require several hours, or days, of
numerical simulation.

\section{Acknowledgments}

\nopagebreak
The authors are grateful to Giuseppe Battistoni, Carlos Castro,
M\'ario Novello, Jane M. Madureira Rached, Nelson Pinto, Alberto Santambrogio, Marisa Ten\'orio de Vasconselos, and particularly
Hugo E. Hern\'andez-Figueroa for many stimulating contacts and discussions. One of us [ER] acknowledges a
past CAPES fellowship c/o UNICAMP/FEEC/DMO.

\

%
%

\end{document}